\documentclass[aps,prstab,twocolumn,showpacs,groupedaddress,floatfix]{revtex4}% for review and submission
\usepackage{graphicx}  % needed for figures
\usepackage{dcolumn}   % needed for some tables
\usepackage{bm}        % for math
\usepackage{amssymb}   % for math
\graphicspath{ {figure/} }
\usepackage{subfig}
\usepackage{float}
\begin{document}

% The following information is for internal review, please remove them for submission
%\widetext
%\leftline{Version xx as of \today}
%\leftline{Primary authors: Joe E. Physics}
%\leftline{To be submitted to (PRL, PRD-RC, PRD, PLB; choose one.)}
%\leftline{Comment to {\tt d0-run2eb-nnn@fnal.gov} by xxx, yyy}
%\centerline{\em D\O\ INTERNAL DOCUMENT -- NOT FOR PUBLIC DISTRIBUTION}

% the following line is for submission, including submission to the arXiv!!
%\hspace{5.2in} \mbox{Fermilab-Pub-04/xxx-E}
\title{Nano-Fabricated Free-Standing Wire Scanners with Sub-Micrometer Resolution}

\author{G.L. Orlandi}
\email{gianluca.orlandi@psi.ch}
\affiliation{Paul Scherrer Institut, 5232 Villigen PSI, Switzerland}
\author{C. David}
\affiliation{Paul Scherrer Institut, 5232 Villigen PSI, Switzerland}
\author{E. Ferrari}
\affiliation{Paul Scherrer Institut, 5232 Villigen PSI, Switzerland}
\author{V. A. Guzenko}
\affiliation{Paul Scherrer Institut, 5232 Villigen PSI, Switzerland}
\author{B. Hermann}
\affiliation{Paul Scherrer Institut, 5232 Villigen PSI, Switzerland}
\affiliation{Institute of Applied Physics, University of Bern, 3012 Bern, Switzerland}
\author{R. Ischebeck}
\affiliation{Paul Scherrer Institut, 5232 Villigen PSI, Switzerland}
\author{E. Prat}
\affiliation{Paul Scherrer Institut, 5232 Villigen PSI, Switzerland}
\author{M. Ferianis}
\affiliation{Elettra-Sincrotrone Trieste S.C.p.A, 34149 Basovizza, Trieste Italy}
\author{G. Penco}
\affiliation{Elettra-Sincrotrone Trieste S.C.p.A, 34149 Basovizza, Trieste Italy}
\author{M. Veronese}
\affiliation{Elettra-Sincrotrone Trieste S.C.p.A, 34149 Basovizza, Trieste Italy}
\author{N. Cefarin}
\affiliation{IOM-CNR Laboratorio TASC, c/o Area Science Park - Basovizza, Trieste, Italy}
\author{S. Dal Zilio}
\affiliation{IOM-CNR Laboratorio TASC, c/o Area Science Park - Basovizza, Trieste, Italy}
\author{M. Lazzarino}
\affiliation{IOM-CNR Laboratorio TASC, c/o Area Science Park - Basovizza, Trieste, Italy}

\date{\today}

\begin{abstract}
Results on fabrication and experimental characterization of wire scanners (WS) with sub-micrometer spatial resolution are presented. Independently fabricated at PSI and FERMI by means of nano-lithography, the proposed WS solutions consist of 900 and 800 $\mathrm{nm}$ wide free-standing stripes ensuring a geometric resolution of about 250 $\mathrm{nm}$. The nano-fabricated WS were tested successfully at SwissFEL where low-charge electron beams with a vertical size of $400-500$ $\mathrm{nm}$ were characterized. Further experimental tests at 200 $\mathrm{pC}$ confirmed the resilience to the heat-loading of the structures. With respect to conventional WS consisting of a metallic wire stretched onto a frame, the nano-fabricated WS allow for an improvement of the spatial resolution from the micrometer to the sub-micrometer level as well as of beam invasiveness thanks to an equivalent reduction of the impact surface of the scanning stripe. The present work represents an important milestone towards the goal --- that PSI and FERMI are pursuing --- to realize a standard WS solution with minimal invasiveness and sub-micrometer resolution.
\end{abstract}

%\pacs{41.60.Cr 41.60.Ap 29.27.Fh}

\maketitle

\section{INTRODUCTION}

Wire scanners (WS) constitute a valuable complement to view screens for monitoring the transverse profile of the electron beam in a linac \cite{fulton,ross,field,tenenbaum,nuhn,loos,wu,wittenburg,hahn,orlandi-WSC-SF}. Because of the multi-shot and mono-dimensional reconstruction of the beam transverse profile, WS are not time-competitive with view screens for beam finding and for matching the magnetic optics in an electron linac. In general, WS are also inappropriate for slice emittance measurements, with some exception. WS measurements of the slice emittance can be indeed performed provided that, in relation to the electron pulse duration, the WS are suitably designed to reconstruct the beam profile from a sufficiently fast readout of the voltage on the wire \cite{huang}. Nevertheless, WS are a unique and essential diagnostics whenever the beam characterization requires a high spatial resolution along with minimal invasiveness to the beam operation. The spatial resolution of a WS depends on the measurement resolution of the wire positioning, on possible wire vibrations and, finally, on the geometry of the wire. The geometric resolution of a WS is inversely proportional to the wire width. This also determines the surface of impact of the wire with the electron beam and hence the wire transparency to the beam (also depending on the wire thickness for non-cylindrical wires). Conventional WS solutions as normally in operation in several free-electron lasers (FELs) are realized according to the standard technique to fix and stretch a metallic wire (beam probe) onto a metallic frame (fork). They are able to attain a spatial resolution at the micrometer level \cite{orlandi-WSC-SF}, which is at least an order of magnitude better than the spatial resolution of a typical view screen operating in an FEL \cite{veronese-bis,ischebeck}. Low-charge and low-emitance machine operation modes --- presently under investigation in several FEL facilities --- requires the characterization of ever smaller beam transverse profiles and, consequently, the development of diagnostics with sub-micrometer resolution. Due to the intrinsic resolution limitation of the conventional WS solutions, new fabrication techniques are to be investigated.

Both PSI and FERMI are independently pursuing the goal of realizing standard WS solutions ensuring sub-micrometer resolution as well as minimal impact on FEL operations. For such a purpose, nano-lithography has been identified by the two laboratories as the most promising fabrication technique. To take the first step at PSI, a nano-lithography WS prototype with sub-micrometer resolution was fabricated and tested on the electron beam \cite{borrelli}. It consisted of a 1 $\mathrm{\mu m}$ wide gold stripe nano-fabricated on-a-membrane ($250$ $\mathrm{nm}$ thick silicon nitride membrane). At FERMI, a WS prototype made of a free-standing $10$ $\mathrm{\mu m}$ wide metallic stripe with a geometric resolution of about $3$ $\mathrm{\mu m}$ was initially nano-fabricated and successfully tested \cite{veronese}.

The work we present here represents a step forward towards the final goal of nano-fabricating an innovative WS solution to be used for standard machine operations in an FEL. The present WS solutions --- nano-fabricated at PSI and FERMI --- join together in the same structure the features of sub-micrometer resolution and free-standing design which, in previous PSI and FERMI prototypes, were achieved separately. They consist indeed of a free-standing scanning stripe fully integrated in a silicon frame.

With stripe widths $(w)$ of $900$ $\mathrm{nm}$ and $800$ $\mathrm{nm}$, respectively, the PSI and FERMI WS attain geometric resolutions ($w/\sqrt{12}$) of about $250$ $\mathrm{nm}$ (see \cite{sigmarms} for the formal definition of the rms geometric resolution in the case of a rectangular wire). They were experimentally tested at SwissFEL both in the nominal- and low-charge regime (bunch charge of $200$ $\mathrm{pC}$ and below $1$ $\mathrm{pC}$, respectively). The electron beam tests demonstrated the capability of these novel WS solutions to resolve beam profiles with a size of $400-500$ $\mathrm{nm}$ as well as the necessary resilience to the heat-loading at nominal machine operations ($200$ $\mathrm{pC}$).

In Sec.~\ref{status-of-art}, we present the state of the art in conventional WS in operation at SwissFEL as well as previous experimental results achieved at PSI and FERMI in the WS nano-fabrication. In Sec.~\ref{nano-fab-tech}, the techniques respectively applied by the two laboratories to nano-fabricate free-standing WS with sub-micrometer resolution are described. In Sec.~\ref{exp-setup-result}, the SwissFEL machine set-up and the results of the electron beam test of the novel sub-micrometer WS are finally reported.

\section{From conventional to nano-fabricated wire scanners}\label{status-of-art}

\subsection{Conventional wire scanners}

At SwissFEL \cite{SwissFEL-CDR, ST, milne}, about $20$ WS are installed all along the machine for routine monitoring of the beam transverse size and for emittance measurements \cite{orlandi-WSC-SF, orlandi-IBIC2017}. The design and fabrication of the SwissFEL WS is based on the traditional technique to fix and stretch thin metallic wires onto a fork. The SwissFEL WS fork is equipped with two independent pairs of cylindrical metallic wires of different material and diameter ($D$): $5$ $\mathrm{\mu m}$ tungsten (W) wires and $12.5$ $\mathrm{\mu m}$ Al(99):Si(1) wires. The spatial resolution of the SwissFEL WS is dominated by the rms geometric resolution of the wires ($D/4$). About the formal definition of the rms geometric resolution of a cylindrical wire, see Ref.~20 in \cite{orlandi-WSC-SF}. With a lower density and a smaller atomic number compared to tungsten, the Al(99):Si(1) wires are dedicated to the routine monitoring of the beam profile during lasing operation at SwissFEL \cite{orlandi-WSC-SF}. The thinner and more precise tungsten wires instead are employed for high-resolution measurements of the emittance when the undulator line is protected by the beam stopper \cite{prat}.

The standard WS manufacture based on a metallic cylindrical wire stretched onto a fork suffers from an intrinsic limitation due to the minimum achievable wire diameter. It is indeed difficult to procure on the market metallic wires with a diameter smaller than a few micrometers. It is even more difficult to find a reliable technique for fixing and stretching such thin metallic wires onto a metallic frame. Hence the necessity to investigate new WS fabrication techniques to improve the geometric resolution of a WS measurement beyond the micrometer scale. The enhancement of the spatial resolution of a WS is also beneficial to the transparency of a WS measurement to the lasing process. With the reduction of the wire width ($w$) and thickness ($t$) --- at the limit $t\rightarrow w$ --- the fraction of electrons sampled by the wire at every machine shot gets smaller as well as the wire induced energy and angular spread of the sampled electrons decrease with benefits for the scanned beam in terms of a better matching with the energy and angular acceptance of the entire machine.

\begin{figure}[H]
%\begin{figure*}
	\centering
	\includegraphics[width=1\linewidth]{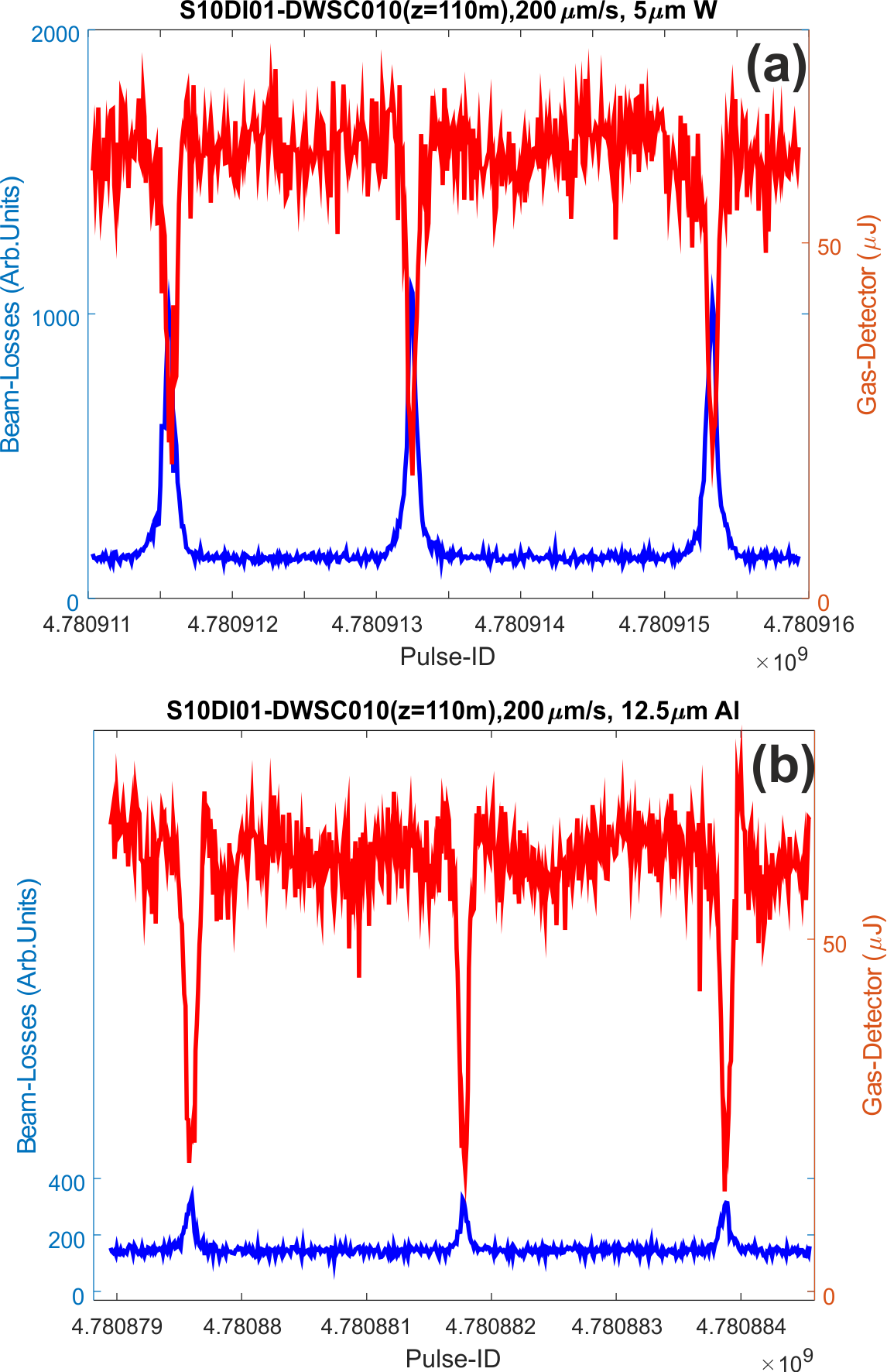}
	\caption{Beam-synchronous measurements of laser pulse energy (gas detector) and electron beam profile (WS) at SwissFEL: (a) $5$ $\mathrm{\mu m}$ W wire; (b) $12.5$ $\mathrm{\mu m}$ Al(99):Si(1) wire. The bunch charge is $200$ $\mathrm{pC}$; the beam energy is $300$ $\mathrm{MeV}$ at the WS location and $2.6$ $\mathrm{GeV}$ at the undulator beamline; photon energy (wavelength) $2.488$ $\mathrm{keV}$ ($4.983$ $\mathrm{{\AA})}$.}
	\label{Orlandi-WS-subum-PRAB-Fig1}
\end{figure}

Spatial resolution, beam invasiveness and lasing transparency are strictly related features of a WS. A comparative study of the wire invasiveness to the beam and transparency to the lasing was carried out at SwissFEL, see Fig.~\ref{Orlandi-WS-subum-PRAB-Fig1}. Measurements of the transverse profile of the electron beam performed with homologous $5$ $\mathrm{\mu m}$ W and $12.5$ $\mathrm{\mu m}$ Al(99):Si(1) wires are shown in Fig.~\ref{Orlandi-WS-subum-PRAB-Fig1} together with beam-synchronous measurements of the FEL pulse energy measured by a gas detector \cite{juranic}. Compared to the W wire, the lower density and smaller atomic number of the Al(99):Si(1) wire are beneficial to the machine protection thanks to a reduction of the beam-losses by a factor $3-4$, as shown in Fig.~\ref{Orlandi-WS-subum-PRAB-Fig1}. On the other hand, the scanning with the $12.5$ $\mathrm{\mu m}$ Al(99):Si(1) wire does not improve the lasing performance because of the larger surface of impact with the beam and the larger number of electrons perturbed by the wire. Finally, when passing from the $5$ $\mathrm{\mu m}$ W to the $12.5$ $\mathrm{\mu m}$ Al(99):Si(1) wire, the observed reduction of the beam losses is consistent with a numerical calculation based on the formula expressing the energy radiated by ultra relativistic electrons in matter \cite{fernow}, see Fig.~\ref{Orlandi-WS-subum-PRAB-Fig1} and Appendix \ref{appendice}.

In conclusion, the room to improve the spatial resolution and the transparency to the lasing of conventional WS made of metallic wires stretched onto a fork seems to be very narrow. These are the premise and the motivation driving PSI and FERMI to investigate new techniques to design and fabricate WS with sub-micrometer spatial resolution.

\subsection{Nano-fabricated wire scanners: first developments}

Electron beam lithography has been chosen at PSI and FERMI as the most promising technique to fabricate minimally invasive wire scanners with sub-micrometer resolution. Thanks to electron beam lithography, it is indeed possible to obtain a complete WS probe and fork system manufactured as a unique structure by applying a nano-fabrication process to an initial lithographic footprint over a substrate.

\begin{figure}[H]
\centering
\subfloat[][]{
\includegraphics[width=0.8\linewidth]{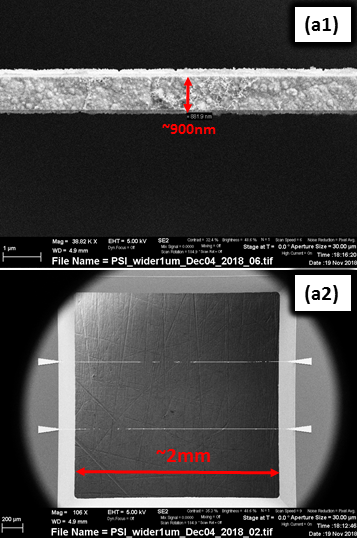}
\label{Orlandi-WS-subum-PRAB-Fig2a}
}\\
\subfloat[][]{
\includegraphics[width=0.8\linewidth]{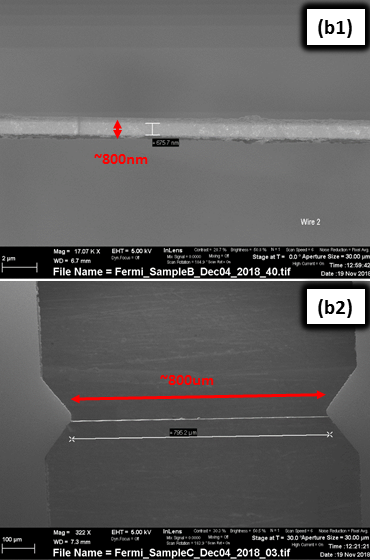}
\label{Orlandi-WS-subum-PRAB-Fig2b}
}
\caption{Scanning Electron Microscope (SEM) images of PSI (a) and FERMI (b) sub-$\mu$m wire-scanners.}
\end{figure}

At PSI, the strategy to achieve this goal passed through the initial nano-fabrication of a WS prototype consisting of a $1$ $\mathrm{\mu m}$ wide Au stripe electroplated onto a thin silicon nitride membrane (WS on-a-membrane). After the successful experimental test at SwissFEL of this prototype \cite{borrelli}, as described in the present work further progress was the nano-fabrication and experimental characterization of a WS consisting of a $2$ $\mathrm{mm}$ long and $900$ $\mathrm{nm}$ wide Au stripe free-standing over a rigid silicon frame, see Fig.~\ref{Orlandi-WS-subum-PRAB-Fig2a}.

At FERMI, the WS nano-fabrication strategy was different. A first WS prototype consisting of a $10$ $\mathrm{\mu m}$ wide Ag/Si$_3$N$_4$/Ag stripe free-standing onto a silicon frame was initially nano-fabricated and tested at FERMI \cite{veronese}. Finally, as described in the following, an upgrade of the previous free-standing WS solution consisting of a $0.8$~$\mathrm{mm}$ long and $800$ $\mathrm{nm}$ wide Au/Si$_3$N$_4$/Au stripe was produced at FERMI and tested at SwissFEL, see Fig.~\ref{Orlandi-WS-subum-PRAB-Fig2b}.

\section{Fabrication of free-standing sub-micrometer wire scanners}\label{nano-fab-tech}

\subsection{Nano-fabrication at PSI}

The fabrication of the free-standing gold nano-wires started with coating a double-side polished Si(001) wafer by a $250$ $\mathrm{nm}$ thick Si$_3$N$_4$ layer. By performing aligned optical lithography on both sides of the wafer, square windows in the optical resist of the size $2\times2$ $\mathrm{mm^2}$ were defined. Applying Reactive Ion Etching (RIE) with CHF$_3$/O$_2$/Ar plasma for $3.5$ minutes, this pattern was transferred into the Si$_3$N$_4$ layer at the top and the bottom of the wafer. Then, the entire top side of the wafer was evaporated by a stack of Ti($5$ $\mathrm{nm}$)/Au($15$ $\mathrm{nm}$)/Ti($5$ $\mathrm{nm}$). The gold layer acts as a seed layer for gold electro-deposition, while titanium improves the adhesion of gold to the wafer or to the e-beam resist. The latter, a film of polymethyl metacrilate (PMMA) with thickness slightly exceeding the target height of the nano-wires by approximatively $100-200$ $\mathrm{nm}$, was spin-coated and baked out on a hot plate at $175$$^{\circ}$C for $15$ minutes. Using a $100$ $\mathrm{keV}$ e-beam lithography system (Vistec EBPG 5000Plus, Raith GmbH), lines of different widths in the range from $300$ $\mathrm{nm}$ to $2$ $\mathrm{\mu m}$, spanning across the Si$_3$N$_4$ windows, i.e., with a length of about $2.5$ $\mathrm{mm}$, were exposed. After the development of the sample in an Isopropanol:DI water ($7:3$ by vol.) solution, a RIE plasma etching step in BCl$_3$ gas was performed (SI500, Sentech GmbH) to remove the upper Ti layer. Subsequently, gold nano-wires were electroplated to the required height (slightly thicker than $2$ $\mathrm{\mu m}$). To turn the fabricated nano-wires into free-standing ones, the metal seed layer as well as the bulk Si had to be removed. First, a sequence of RIE with BCl$_3$ and the Ar ion beam milling was performed until reaching the Si surface. The very last step of removing Si inside the Si$_3$N$_4$ window supporting the nano-wires was completed in a KOH bath (20$\%$ wt. in H$_2$O) at $75$$^{\circ}$C. Finally, the samples with free-standing nano-wires were rinsed in a hot DI H$_2$O and slowly dried in air.

\begin{figure}[H]
%\begin{figure*}
	\centering
	\includegraphics[width=1\linewidth]{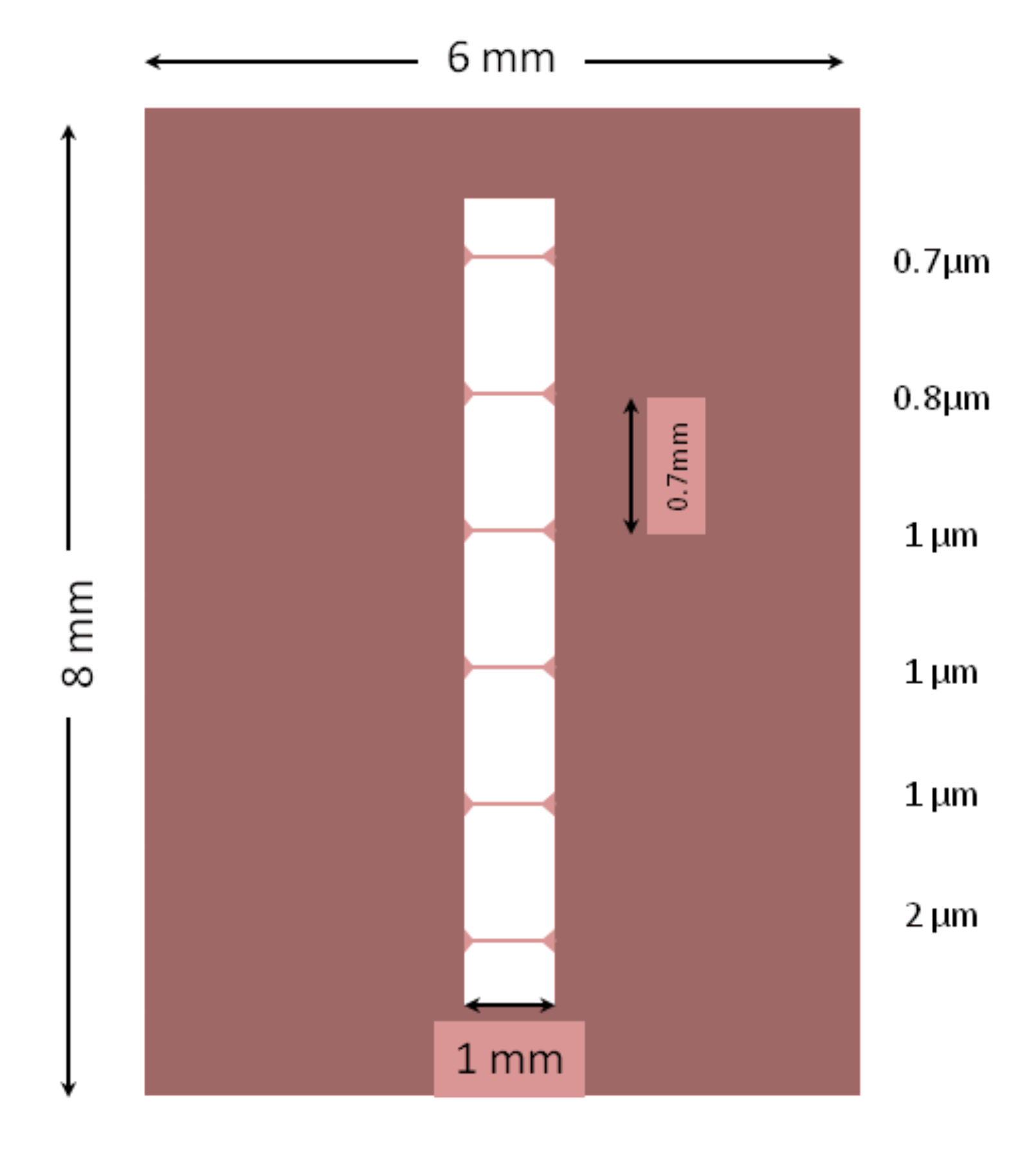}
	\caption{Schematic drawing of the FERMI free-standing WS chip with indication of the width of the free-standing stripes allocated on it.}
	\label{Orlandi-WS-subum-PRAB-Fig3}
\end{figure}
%\end{figure*}

\begin{table*}[t]
\caption{\label{Tabella1} Analysis results of the measurements of the vertical size of the SwissFEL electron beam by means of the PSI and FERMI free-standing WS in the measurement sessions of December 4, 2018 and March 31, 2019. Experimental data analysis performed by means of the error-function based fitting function, see Eq.~(\ref{eq1}). Machine setting: beam charge less than $1$ $\mathrm{pC}$; beam energy $300$ $\mathrm{MeV}$; vertical emittance about $55$ $\mathrm{nm}$; expected vertical beam size at the WS position of about $480$ $\mathrm{nm}$ for a beta function value $\beta_y=2.61\times10^{-3}$ $\mathrm{m}$.}

\begin{ruledtabular}
\begin{tabular}{ccccc}
%\begin{tabular}{rrrrr}
%\begin{tabular}{lllll}
\textbf{WS type}&
\textbf{stripe width($\mathrm{nm}$)}&
\textbf{geom. res.($\mathrm{nm}$)}&
\textbf{beam size ($\mathrm{nm}$, Dec 2018)}&
\textbf{beam size ($\mathrm{nm}$, Mar 2019)}\\
\colrule
\textbf{PSI-WS} & $900$ & 260 & 488$\pm$20 & 434$\pm$7\\
\textbf{FERMI-WS} & $800$ & 230 & 477$\pm$70 & 443$\pm$33\\
\end{tabular}
\end{ruledtabular}
\end{table*}

\subsection{Nano-fabrication at CNR-IOM}

\begin{figure}[H]
%\begin{figure*}
	\centering
	\includegraphics[width=1\linewidth]{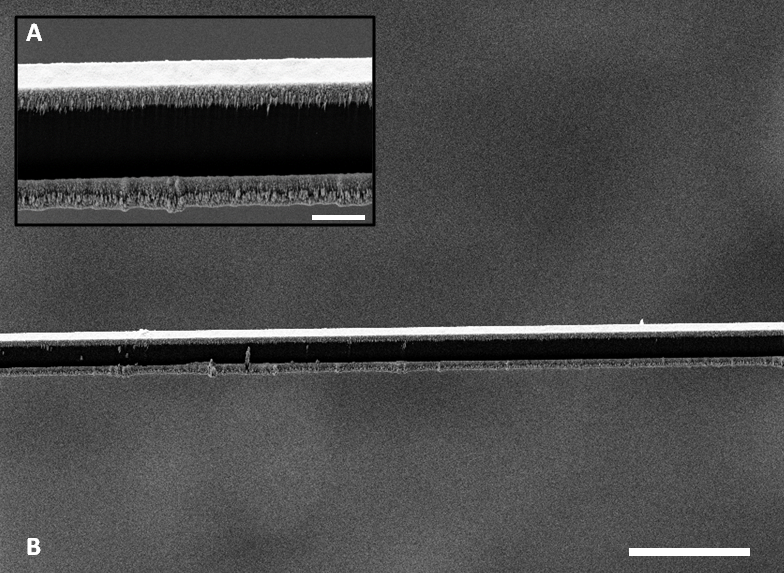}
\caption{SEM image of the Si$_3$N$_4$ micro wire. (A) Example of produced wires, after the metallization (scale bar 10 $\mathrm{\mu m}$). (B) A detail of the structure: visible are the gold depositions on both the top and bottom parts of the wire, consisting of a thin TiAu bilayer (10-500 $\mathrm{nm}$). The deposition parameters were set in order to reduce as much as possible the presence of gold on the lateral side of the wire (scale bar 1 $\mathrm{\mu m}$).}
%\caption{SEM image of a detail of the Si$_3$N$_4$ wire. The brighter part is the Au coating, while the darker region is the Si$_3$N$_4$ suspended wire. The deposition parameters were set in order to reduce as much as possible the presence of gold on the lateral side of the wire.}
	\label{Orlandi-WS-subum-PRAB-Fig4}
\end{figure}

Low stress silicon nitride (SiN) is one of the best candidate for the production of high-resolution suspended structures thanks to its mechanical properties. Currently employed as structural material in MEMS devices, it permits to produce extreme aspect ratio structures, such as large membranes or long cantilevers. Also the wire employed as scanner in FEL applications can be considered in the same way as a long cantilever. The idea is to exploit the MEMS technology derived from semiconductor industry, so ensuring high flexibility in terms of fabrication process and device capabilities. In Fig.~\ref{Orlandi-WS-subum-PRAB-Fig3} it is shown schematically the proposed geometry: the device hosts a series of wires with different lateral thicknesses: from $\mathrm{\mu m}$ microns up to sub-micrometer size, just achievable with the proposed technology.
Nonetheless, the fragility of the proposed structures imposes fine control of the tensile residual film stress inherent to the deposition process, which deeply affects the functionality and reliability of MEMS devices.  Internal mechanical stresses indeed induce warps of free-standing micro-structures (buckle up, bent down or ultimately cracks) causing complete failure. The use of low-stress LPCVD silicon nitride ($<250$ MPa tensile stress, from SI-MAt), such as the fabrication process design, allows preventing or reducing these effects to increase the reliability and the quality of the final devices.
The fabrication procedure employed a silicon wafer coated with 2 $\mathrm{\mu m}$ low-stress LPCVD silicon nitride. After the standard cleaning procedure in RCA-1 solution (H$_2$O$_2$:NH$_4$OH:H$_2$O 1:1:5, $60$$^{\circ}$C), 150 $\mathrm{nm}$ thick chromium film was deposited by DC magnetron sputtering on both sides. On the back side, the pass-trough area of the device was defined by classical proximity UV lithography. After the wet etching of the exposed chromium layer, the SiN film was etched in ICP-RIE (SF$_6$-O$_2$-C$_4$F$_8$, power), and the silicon was thinned up to 50 $\mathrm{\mu m}$ in KOH solution (33\% in wt, $80$$^{\circ}$C). The silicon was not entirely removed at this stage to make the structure to withstand the following processes.
A second UV lithography/Cr etch step allowed defining the window where the wires were then hosted on the front side. Before dry etching the SiN layer, the pattern of the micro wire was defined by exposing a second layer of resist (PMMA AR-P671-05, 2000 rpm) by an EBL like system (Zeiss Leo 1540 equipped with external module Raith pattern generator): on each device we produced six structures with different width. After the deposition of a $100$ $\mathrm{nm}$ thick Ni mask on the samples, the structures were transferred in the SiN by lift-off followed by dry etching in ICP-RIE. The samples were then transferred in KOH for the removal of the residual layer of silicon on the bottom of the wires; finally the device was rinsed from KOH by a suitable procedure, consisting in the sequential dip into hot water cold water and in water solutions with increasing content of ethanol, till the dipping in pure ethanol. The drying by low evaporation of solvent in ethanol saturated ambient avoided the suspended structures to collapse under the surface tension forces of the solvent. The slow drying was also tested from pure water, but independently of the drying time, we always observed the brake up of the wires. During the whole procedure the temperature was always slowly varied at a rate below $1$$^{\circ}$C/min to avoid the creation of sudden thermal stress that may damage the structure.
The devices were then transferred in RIE for the cleaning from the ethanol residuals in O$_2$ plasma. The last process step was the deposition of Ti-Au ($10-500$ $\mathrm{nm}$) thin film by e-gun evaporation. The deposition rate was maintained as low as 0.5 $\mathrm{nm}$/$\mathrm{s}$ to avoid excessive heating or damage of the structures. In particular we observed the need for depositing on both sides the metallic coating: this is important to prevent undesired wire bowing. An example of the achieved structures is shown in Fig.~\ref{Orlandi-WS-subum-PRAB-Fig4}. It is worth noting that the metal film provides also good electrical conductivity to the whole structure, thus avoiding possible charging effects during the FEL beam exposure: SiN is an insulating material, and the charging of the structures generated by the interaction with the beam cannot be drained off without a metallic coating.

%\end{figure*}

\begin{figure}[H]
%\begin{figure*}
	\centering
	\includegraphics[width=0.8\linewidth]{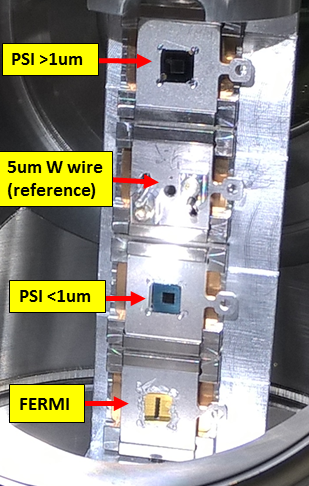}
	\caption{Free-standing WS with sub-micrometer resolution installed in the sample holder. From top to bottom: PSI chip equipped with $2$ Au stripes with a width of $2$ and $1$ $\mathrm{\mu m}$ (PSI~$>1$ $\mathrm{\mu m}$); PSI chip equipped with a standard cylindrical W wire with a diameter of $5$ $\mathrm{\mu m}$ ($5um-W-wire$); PSI chip equipped with $2$ Au stripes with a width of $1$ and $0.9$ $\mathrm{\mu m}$ (PSI~$<1$ $\mathrm{\mu m}$); FERMI chip equipped with $6$ Au stripes width ranging from $2.5$ to $0.7$ $\mathrm{\mu m}$ (FERMI). As described in the present work, the electron-beam characterization at SwissFEL mainly focused on the free-standing WS with the best geometric resolution, i.e., the PSI $0.9$ $\mathrm{\mu m}$ and FERMI $0.8$ $\mathrm{\mu m}$ wide stripes.}
	\label{Orlandi-WS-subum-PRAB-Fig5}
\end{figure}
%\end{figure*}

\begin{figure*}
    \includegraphics[width=1\linewidth]{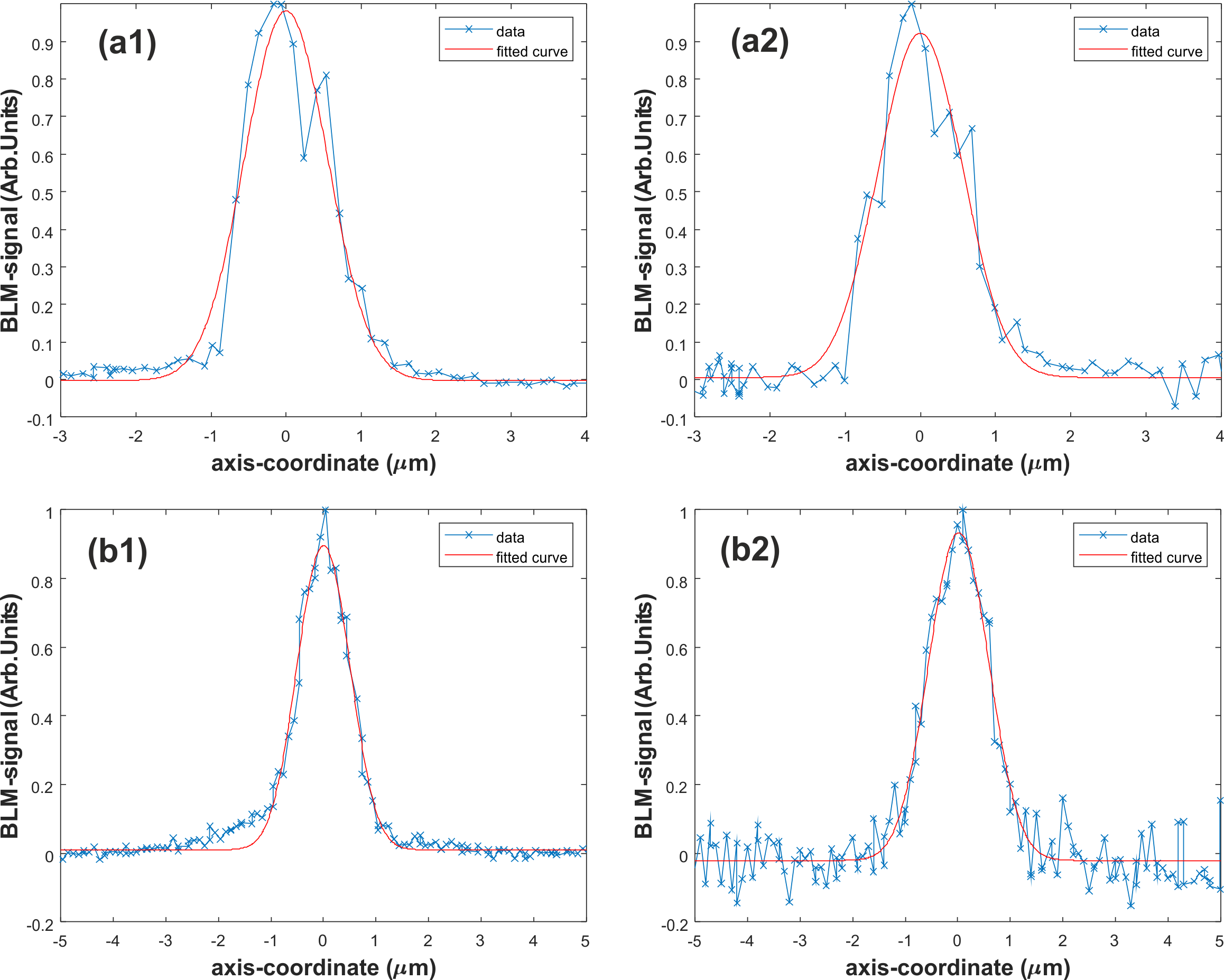}
    \caption{Beam vertical profiles measured by means of the free-standing WS with sub-micrometer resolution: PSI $900$ $\mathrm{nm}$ (a1, b1) and FERMI $800$ $\mathrm{nm}$ (a2, b2) wide Au stripe. Measurements have been performed at SwissFEL in two different machine sessions: (a) December 4, 2018; (b) March 31, 2019. The beam profile plotted in pictures (a) is the result of the average of 5 acquisitions, while the beam profiles plotted in pictures (b) are the result of a single shot acquisition. Beam size determined by means of error-function fit, see Eq.~(\ref{eq1}) and Tab.~\ref{Tabella1}.}
    \label{Orlandi-WS-subum-PRAB-Fig6}
\end{figure*}

\section{Experimental set-up and results}\label{exp-setup-result}

The electron-beam test of both the PSI and FERMI prototypes of free-standing WS with sub-micrometer resolution was carried out in parallel at SwissFEL during several measurement sessions.

SwissFEL is an X-ray FEL facility \cite{SwissFEL-CDR, ST, milne} in operation at the Paul Scherrer Institut. Driven by an rf linac --- a S-band injector and a C-band accelerator --- in the beam energy range $2.1-5.8$ $\mathrm{GeV}$, SwissFEL is presently producing tunable and coherent hard X-ray pulses in the wavelength region $0.7-0.1$ $\mathrm{nm}$ (Aramis undulator line) and, by 2021, will also generate soft X-ray radiation in the wavelength region $7-0.7$ $\mathrm{nm}$ (Athos undulator line).

The characterization of the free-standing WS was performed at a beam energy of about $300$ $\mathrm{MeV}$ downstream of the SwissFEL injector in a multi-functional vacuum chamber, the so called ACHIP chamber \cite{ferrari}. In the ACHIP chamber, a $4$-slot sample holder can be inserted vertically into the beam line by means of a UHV feed-through motorized by a 2-phase stepper motor and equipped with an encoder. Therefore, only WS measurements of the vertical profile of the electron beam are possible. Thanks to a load-lock pre-vacuum chamber, the WS structures can be positioned onto the sample holder of the chamber with minimal perturbation of the machine vacuum. The picture in Fig.~\ref{Orlandi-WS-subum-PRAB-Fig5} shows the FERMI nano-fabricated WS (at the very bottom of the sample holder), the 2 PSI nano-fabricated WS and, in between them, a reference WS consisting of a $5$ $\mathrm{\mu m}$ W wire. At SwissFEL, a WS measurement of the beam transverse profile is achieved by correlating the wire position measured by the encoder and the beam-losses detected by a beam-loss monitor (BLM) \cite{cigdem} that, in this particular case, is about $2$ $\mathrm{m}$ downstream of the vacuum chamber. Studies to optimize the distance between WS and BLM at SwissFEL are reported in \cite{orlandi-WSC-SF}.

The PSI and FERMI free-standing WS were tested in parallel with the electron beam of SwissFEL on two different days at a beam energy of about $300$ $\mathrm{MeV}$. In both measurement sessions, SwissFEL was set in a low-emittance and low-beta mode \cite{borrelli}. Low-emittance operations at SwissFEL are possible thanks to a low-charge setup (below $1$ $\mathrm{pC}$) of the rf gun photo-cathode, which allows for a normalized vertical emittance ($\varepsilon_{n,y}$) of about $55$ $\mathrm{nm}$. Thanks to a suitable beam optics (low-beta), at the wire interaction point the beta functions ($\beta_x,\beta_y$) are ($0.27,2.61\times10^{-3}$) $\mathrm{m}$. At the WS position, the vertical beam size $(\sigma_y=\sqrt{\beta_y\varepsilon_{n,y}/\gamma})$ is hence about $400-500$ $\mathrm{nm}$ where $\gamma$ is the relativistic Lorentz factor at a beam energy of $300$ $\mathrm{MeV}$. The beam horizontal size is about 10 times larger.

In Fig.~\ref{Orlandi-WS-subum-PRAB-Fig6}, several beam vertical profiles acquired at SwissFEL with the PSI $900$ $\mathrm{nm}$ and the FERMI $800$ $\mathrm{nm}$ free-standing WS are shown together with a fitting function. To take into account the finite size of the wire width in the analysis of the experimental data, the applied fit function is the convolution of a Gaussian distribution and a rectangular shaped distribution modelling the cross section of the WS stripe. In this way, the measured beam size resulting from the analysis of the beam profile is not affected by the finite geometric resolution of the stripe provided that the electron beam size is at least of the order of the resolution limit. Under the assumption that the beam profile can be approximated reasonably well by a Gaussian distribution with standard deviation $\sigma$, the convolution of the beam distribution with the rectangular distribution modelling the WS stripe leads to a formal expression that depends on the error function (erf) and reads as
\begin{eqnarray}
	f(x) =&& a\times[\mathrm{erf}([x-c+w/2]/\sqrt{2}/\sigma)+\nonumber\\
&&-\mathrm{erf}([x-c-w/2]/\sqrt{2}/\sigma)]+b,\label{eq1}
\end{eqnarray}
where $a$, $b$, $c$ and $\sigma$ are free parameters of the fit function ($a$ accounting for the signal amplitude, $b$ for the off-set and $c$ for the centroid coordinates of the Gaussian and rectangular distribution functions) and $w$ is a predefined constant parameter accounting for the stripe width.

%\begin{figure}[H]
\begin{figure}[H]
%\begin{figure*}
	\centering
	\includegraphics[width=1\linewidth]{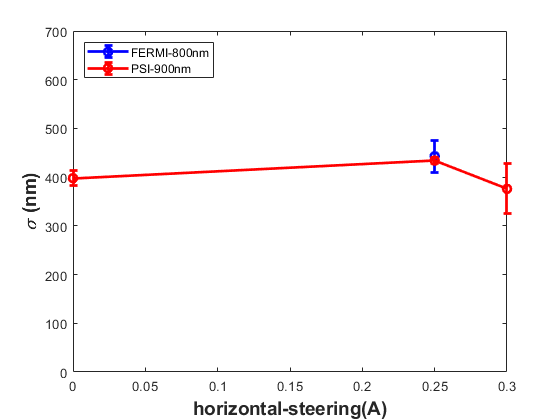}
	\caption{PSI and FERMI WS measurements of the beam vertical size $\sigma$ obtained in the machine session of March 31, 2019. Beam size measurements obtained by steering horizontally the electron beam in the vacuum chamber, while keeping constant the machine optics, i.e, without changing the beam transverse size.}\label{Orlandi-WS-subum-PRAB-Fig7}
\end{figure}
%\end{figure*}

In Tab.~\ref{Tabella1} the comparison of the vertical size of the beam profile measured in the two distinct measurement sessions is reported. The measurements are the result of the average of five consecutive acquisitions. In both measurement sessions the beam transverse profile measured by the PSI and FERMI WS are consistent within statistical errors.

A different signal-to-noise ratio characterizes the PSI and FERMI WS measurements as it is evident in the normalized beam profiles plotted in Fig.~\ref{Orlandi-WS-subum-PRAB-Fig6}. Further evidence for a difference in signal-to-noise ratio between the PSI and FERMI WS measurements can be found in the statistical error of the beam size, which is much larger in the FERMI case compared to the PSI one, see Tab.~\ref{Tabella1}. As already described in Sec.~\ref{nano-fab-tech}, this difference is due to the different fabrication techniques of the WS structures. The FERMI WS stripe being a sandwich structure Au/Si$_3$N$_4$/Au (Au total thickness about $1$ $\mathrm{\mu m}$) is characterized by a radiation length \cite{fernow} about two times longer than the PSI WS which instead consists of a bulk Au stripe (Au total thickness about $2$ $\mathrm{\mu m}$). The FERMI WS is less invasive (i.e., less dose released), therefore resulting in a noisier signal. The best Au layer thickness has to be a trade-off between a proper signal-to-noise ratio and the tolerable released dose.

WS measurements of the vertical beam size $\sigma $ for different values of the horizontal position of the beam centroid are shown in Fig.~(\ref{Orlandi-WS-subum-PRAB-Fig7}). They were obtained by steering horizontally the electron beam in the vacuum chamber still keeping constant the machine optics, i.e, without changing the beam transverse size. The consistent agreement within the statistical errors of the measurements in Fig.~(\ref{Orlandi-WS-subum-PRAB-Fig7}) constitutes a check of the measurement reproducibility as a function of the horizontal position of the beam centroid along the WS stripe as well as a further confirmation of the validity of the experimental procedure described in the present work.

We also performed a resilience test of the free-standing WS to heat-loading at high-charge. The test was carried out at the nominal high-charge operation mode of SwissFEL ($200$ $\mathrm{pC}$). No damage due to the heat-loading was observed in the free-standing WS after several and repeated series of measurements at a beam charge of $200$ $\mathrm{pC}$.

In conclusion, with reference to the data reported in Tab.~\ref{Tabella1} and in Fig.~\ref{Orlandi-WS-subum-PRAB-Fig6}, a beam transverse profile of less than $500$ $\mathrm{nm}$ was consistently measured by both the PSI and FERMI free-standing sub-micrometer WS in two distinct measurement sessions at SwissFEL.

\section{Conclusions and Outlook}

PSI and FERMI are independently pursuing research and development programs aiming at improving the spatial resolution of wire scanners (WS) beyond the standard limit of the micrometer scale as well as the WS transparency to the lasing operations in an FEL. Nano-lithography permits to overcome the bottleneck of the micrometer resolution limit, which characterizes the conventional WS design consisting of a metallic wire stretched over a metallic fork. It is indeed possible to nano-fabricate free-standing Au bulk or sandwich Au/Si$_3$N$_4$/Au WS stripes with sub-micrometer width, which are fully integrated into a silicon frame. In the present work, we reported on the production details of two different prototype solutions of free-standing WS independently nano-fabricated at PSI and FERMI and experimentally tested at SwissFEL. With a $900$ and $800$ $\mathrm{nm}$ wide scanning stripe directly nano-fabricated onto a silicon frame, the PSI and FERMI free-standing WS achieve a geometric resolution of about $250$ $\mathrm{nm}$, respectively. The PSI WS prototype consists of a $2$ $\mathrm{mm}$ long bulk Au stripe, while the FERMI prototype consists of a $0.8$ $\mathrm{mm}$ long stripe made of a sandwich of Au/Si$_3$N$_4$/Au.
Both WS prototypes were tested in parallel at SwissFEL under a low-charge and low-emittance setting of the machine as well as under the nominal high charge mode of the machine ($200$ $\mathrm{pC}$) to check the resilience of the structures to the heat-loading. Under a low emittance setting of the machine, at the interaction point the electron beam reached a vertical size of $400-500$ $\mathrm{nm}$, which was consistently measured within the statistical errors by the two WS prototypes in two different experimental sessions. The data analysis being performed by means of a Gaussian fit function convoluted with a rectangular distribution modelling the cross section of the wire permitted to distinguish and separate in the measured beam profile the contribution due to the beam size from the one due to the finite geometric resolution of the scanning stripe. At present, the nano-fabricated free-standing WS solutions can ensure a beam clearance of about $2$ $\mathrm{mm}$. For a routine use of the nano-fabricated WS as a standard solution in a linac driven FEL, the present beam clearance should be increased by a factor $4-5$, at least. This design improvement is in the to-do list of the development program of nano-fabricated free-standing WS with sub-micrometer resolution at PSI and FERMI.

In conclusion, the described campaign of WS measurements --- carried out at SwissFEL in two different beam sessions --- confirmed the expected high resolution performance of the PSI and FERMI free-standing WS in resolving electron beam profile with sub-micrometer size and demonstrated the reliability of two different techniques of nano-fabrication independently implemented at PSI and FERMI. The way to the implementation of nano-fabricated WS with sub-micrometer resolution as a standard WS solution in FELs is paved.

\section{Acknowledgements}
The authors wish to thank the Paul Scherrer Institut expert groups, the SwissFEL commissioning and operation team for the support during the measurements.
The authors also wish to thank S. Borrelli, O. Huerzeler, C. Lombosi and C. Ozkan-Loch for useful comments and support. The authors are grateful to P. Jurani\'c for his support with the gas-detector measurements. For the careful proofreading of the manuscript and fruitful comments, the authors wish to thank T. Schietinger. This work was also supported by the Gordon and Betty Moore Foundation (ACHIP collaboration).

\appendix

\section{}\label{appendice}

The reduction by a factor $3-4$ of the beam-losses observable in Fig.~\ref{Orlandi-WS-subum-PRAB-Fig1} --- when passing from the $5$ $\mathrm{\mu m}$ W to $12.5$ $\mathrm{\mu m}$ Al(99):Si(1) wire --- is consistent with a numerical calculation based on the formula expressing the energy radiated by an ultra relativistic electron in matter \cite{fernow}:
\begin{eqnarray}
\frac{dE}{E}=\frac{dX}{L_R},\label{eq2}
\end{eqnarray}
where $dE$ is the infinitesimal energy radiated by an electron of energy $E$ when crossing an infinitesimal thickness $dX$ of a material with a radiation length $L_R$. With reference to Eq.~(\ref{eq2}), by taking into account the ratio of the impact surface $R_{W/Al}=0.4$ of the W wire with respect to the Al(99):Si(1) wire, the relative ratio of the totally radiated energy by the electron beam in the two cases reads
\begin{eqnarray}
\frac{\Delta E_W}{\Delta E_{Al}}=R_{W/Al}\frac{X_W}{L_W}\frac{L_{Al}}{X_{Al}}=4.1,\label{eq3}
\end{eqnarray}
where $X_W$ and $X_{Al}$ are the rms diameters of the wires, $L_W=0.35$ $\mathrm{cm}$ and $L_{Al}=8.9$ $\mathrm{cm}$ are the radiation length of tungsten and aluminium, respectively.

\end{document}